\pdfoutput=1
\documentclass[
 nofootinbib,
 amsmath,amssymb,
 aps,
 prd,
]{revtex4-1}
\usepackage[utf8]{inputenc}

\usepackage{graphicx}
\usepackage{dcolumn}
\usepackage{bm}
\usepackage{xcolor}
\usepackage{enumitem}
\usepackage{printlen}
\usepackage{braket}
\usepackage{amsmath}
\usepackage{booktabs}
\usepackage{array}
\usepackage{verbatim}

\usepackage{natbib}

\DeclareGraphicsExtensions{.pdf,.png}

\newcommand{\dd}[1]{\mathrm{d}{#1} \:}
\newcommand{\Msun}{\ensuremath{\mathrm{M}_\odot}}

\newcommand{\nobs}  {{\ensuremath{N_\mathrm{obs}}}}
\newcommand{\nobsfg}{\ensuremath{N_\mathrm{F,obs}}}
\newcommand{\nobsbg}{\ensuremath{N_\mathrm{B,obs}}}
\newcommand{\nexp}  {\ensuremath{N_\mathrm{exp}}}
\newcommand{\nexpfg}{\ensuremath{N_\mathrm{F,exp}}}
\newcommand{\nexpbg}{\ensuremath{N_\mathrm{B,exp}}}
\newcommand{\popfg} {\ensuremath{\theta_\mathrm{F}}}
\newcommand{\popbg} {\ensuremath{\theta_\mathrm{B}}}
\newcommand{\class} {\ensuremath{\eta_i}}
\newcommand{\vclass}{\ensuremath{\{\eta}\}}
\newcommand{\isfg}  {\ensuremath{\eta_i \! = \! \mathrm{F}}}
\newcommand{\isbg}  {\ensuremath{\eta_i \! = \! \mathrm{B}}}
\newcommand{\snr}   {\ensuremath{\rho_i}}
\newcommand{\vsnr}  {\ensuremath{\{\rho}\}}
\newcommand{\PE}  {\ensuremath{\vec{\gamma}_i}}
\newcommand{\vPE} {\ensuremath{\{\vec{\gamma}\}}}
\newcommand{\rate}  {\ensuremath{R}}

\newcommand{\obssnr}   {{\ensuremath{\rho_\mathrm{obs}}}}
\newcommand{\truesnr}  {{\ensuremath{\rho_\mathrm{true}}}}
\newcommand{\threshold}{{\ensuremath{\rho_\mathrm{thr}}}}
\newcommand{\cutoff}   {{\ensuremath{\rho_\mathrm{cutoff}}}}

\begin{document}

\preprint{LIGO-TXXXXXXX}

\title{Digging the population of compact binary mergers out of the noise}

\author{Sebastian M. Gaebel}
\affiliation{Institute for Gravitational Wave Astronomy, School of Physics and Astronomy, University of Birmingham, Edgbaston, Birmingham, B15 2TT, United Kingdom}

\author{John Veitch}
\affiliation{Institute for Gravitational Research, School of Physics and Astronomy, University of Glasgow, Glasgow, G12 8QQ, United Kingdom}

\author{Thomas Dent}
\affiliation{Max Planck Institute for Gravitational Physics (Albert Einstein Institute), D-30167 Hannover, Germany}

\author{Will M. Farr}
\affiliation{Institute for Gravitational Wave Astronomy, School of Physics and Astronomy, University of Birmingham, Edgbaston, Birmingham, B15 2TT, United Kingdom}
\affiliation{Center for Computational Astrophysics, Flatiron Institute, 162 Fifth Avenue, New York NY 10010, USA}
\affiliation{Department of Physics and Astronomy, Stony Brook University, Stony Brook, NY, 11794, USA}

\date{\today}

\begin{abstract}
Coalescing compact binaries emitting gravitational wave (GW) signals, as recently detected by the Advanced LIGO-Virgo network, constitute a population over the multi-dimensional space of component masses and spins, redshift, and other parameters.  Characterizing this population is a major goal of GW observations and may be approached via parametric models.  We demonstrate hierarchical inference for such models with a method that accounts for uncertainties in each binary merger's individual parameters, for mass-dependent selection effects, and also for the presence of a second population of candidate events caused by detector noise.  Thus, the method is robust to potential biases from a contaminated sample and allows us to extract information from events that have a relatively small probability of astrophysical origin. 
\end{abstract}

\maketitle

\section{Introduction}

Since the first detection of gravitational waves in 2015~\cite{GW150914}, the Advanced LIGO and
Advanced Virgo detectors have observed the coalescence of multiple compact binary systems,
and have begun to reveal the population of coalescing compact objects~\cite{O1BBH}.
This population is enabling studies in fields from probing alternative theories
of gravity to constraining models of stellar evolution. These tend to be interested
either in individual, preferably loud, signals, or in the population of sources as a whole.
The latter type of population analysis tries to estimate the parameters governing the distribution of 
sources in the universe, their masses and spins, and the value of the astrophysical merger rate~\cite{GW150914:rates}
are of particular interest. As the sensitivity of detectors improves over the coming
years, the detected number of sources is expected to grow at an accelerated pace, rapidly
increasing the amount of information available for population studies~\cite{ObservingScenarios}.


When undertaking population analyses, one has to consider that real detectors may 
produce noise transients which cannot in all cases be distinguished from astrophysical 
GW sources.  To avoid population inferences being biased by noise events one might 
consider only events with a much higher probability to be of astrophysical origin than 
to be caused by noise artefacts. 
In templated searches for compact binaries, the relative probability of astrophysical
vs.\ noise origin for a candidate event is a function of a detection statistic calculated
for each event by a search analysis pipeline (see e.g.~\cite{GW150914:CBCsearch,Usman:2015kfa,Cannon:2011vi,
2015arXiv150404632C,Messick:2016aqy,Nitz:2017svb}).  
Typically a candidate event is generated as a local maximum in matched filter 
signal-to-noise ratio (SNR) above a search threshold; the detection statistic value 
then incorporates the matched filter SNR, as well as other goodness-of-fit tests to 
reject non-Gaussian instrumental noise transients~\cite{Allen:2004gu,Nitz:2017lco}. 

At low SNR, the population of events is dominated by the noise `background', 
whereas at high SNR (or in general for events assigned high statistic values) 
the astrophysical `foreground' dominates.\footnote{We will loosely refer to the 
detection statistic as `SNR' when discussing the distinction between instrumental
noise events and astrophysical events.} 
To limit possible pollution of the sample used for population inference, one may 
place a minimum threshold on the detection statistic; any event above threshold 
is then assumed to be astrophysical, whereas all other events are discarded as 
potential noise transients. 
Note that the choice of threshold value requires an empirical estimate of the 
rate and distribution of background events~\cite{Capano:2016uif}, since the 
rate, strength and morphologies of detector noise artefacts are not known 
\emph{a priori}~\cite{TheLIGOScientific:2016zmo}. 

A simple strategy of thresholding is sub-optimal for two reasons.  First, discarding 
events below the threshold will almost certainly discard information from some number
of quiet but still identifiable signals; second, there is still a finite chance that 
the resulting `signal' set is nevertheless contaminated by noise, leading to potentially
biased inferences.  The choice 
of SNR threshold requires a trade-off between these two considerations and depends 
on the intended use. One also has to take into consideration any bias in the observed
population produced by the effect of source parameters on the loudness of a signal, 
and thus its chance of exceeding a SNR threshold~\cite{Fairhurst:2007qj}; we expect 
potentially major observation selection effects for binary mass(es)~\cite{Tiwari:2017ndi} 
and component spins~\cite{Ng:2018neg}.

Here, we propose a method which alleviates the issues associated with simple 
thresholding by applying a hierarchical mixture model under which each event is 
considered to originate from either a foreground or a background (noise) population.  For
each event, the probability of either case naturally defines a weight for its 
contribution to inferences on population parameters. 

This method combines the processes of estimating the expected number of events 
in either class~\cite{FGMC} (the number of foreground events being a proxy for the 
astrophysical merger rate density) and estimating parameters of the underlying
populations, which have previously been performed separately. 
It avoids being biased through the inclusion of background events,
while being able to make use of events with a non-negligible probability of noise
origin, which would be discarded by thresholding.  In theory this method 
allows the SNR threshold to be reduced to an arbitrarily low value, though in 
practice we are still limited by the computational resources required to extract
the source parameters from each event under consideration.\footnote{An analysis 
that effectively removes all SNR thresholds, applying Bayesian analysis to the 
entirety of the gravitational-wave data set rather than restricting to data close
to events triggered on SNR maxima, is proposed in~\cite{Smith:2017vfk}; its 
application appears at present to be still more limited by computational cost.}

Our method is applicable to any hierarchical model of a source population,
examples of which have been explored in the literature.
This includes analyses which combine information
from multiple events to infer a parameter common to all, such as deviations
from general relativity~\cite{Li:2011cg} or a parameterised neutron star equation of
state~\cite{DelPozzo:2013ala}.
The use of a mixture model with astrophysical and noise populations is particularly
useful when the model of interest has a strong effect on the detectability of
sources, i.e.\ the detected events are unrepresentative of the underlying population.
A good example is the mass distribution of sources, which we consider below.

\citet{Messenger:2012jy} considered the problem of selection effects in mass
distribution inference by dividing the observing time into discrete chunks, which
each contain zero or one sources, and computing population likelihoods while accounting
for false alarms and false dismissals from an idealised noise distribution.
\citet{FGMC} derived an equivalent formalism for rate inference that allows
a population shape function to be estimated alongside. Our derivation in 
Section~\ref{sec:notation_and_method} follows similar lines.

The selection function for masses was important in estimating the astrophysical
event rates in the first Advanced LIGO observing run (O1), which inferred rates
using a mixture model, for fixed choices of population shape (i.e.\ mass distribution)~\cite{GW150914:rates,GW150914:rates_supplement}. 
A separate analysis also estimated the slope of a power law model of the mass 
distribution function considering the detected events, described in~\cite{O1BBH} 
(and updated in~\cite{GW170104}). 

The selection function in the form of a sensitivity-weighted measure of the space-time 
volume $VT$ is also important when considering searches which do not make a clear detection.
There, the loudest background event, or a nominal detection threshold, is used to set 
an upper limit on the astrophysical rate of a fiducial source population, for example
limits on the rate of mergers of binary neutron stars and neutron star--black hole binaries in
O1~\cite{O1:BNSandNSBH}. 
The sensitive-volume approach has been in use since the initial detector 
era~\cite{Biswas:2007ni,S6upperlimits}, and continues to be refined to incorporate 
mass- and spin-dependent selection effects~\cite{PhysRevD.82.104006,
2015ApJ...806..263D, Ng:2018neg} and cosmological effects, as well as
to improve the accuracy of measurement~\cite{Tiwari:2017ndi}. 

As the number of detections increases, determination of the population of
coalescing compact binaries is expected to provide insight into the astrophysics
of black hole and neutron star binary formation, as reviewed 
in~\cite{kalogera2007,gw150914astrointerpretation,Mandel:2018hfr}).
Population synthesis models can describe the masses and spins of coalescing compact 
binaries under a variety of formation scenarios (see e.g.~\cite{Belczynski:2016jno,
2017arXiv170607053B,2016Natur.534..512B,Spera:2015vkd}). 
Comparison of these predictions to the observed distribution can be used to
constrain the uncertainties in parametrised models of source populations~\cite{Barrett:2017fcw,Zevin:2017evb}. 
This has motivated the development of methods to determine the mass-dependent
coalescence rate in the absence of false alarms, using both specific parameterised
models~\cite{Wysocki:2017isg,Wysocki:2018mpo,2018ApJ...856..173T,TaylorGerosa:2018}
and non-parametric methods~\cite{Mandel:2016prl}. 
The alignment of black hole spins is expected to be a key distinguishing feature between
binaries formed in the field or through dynamical interactions (see e.g.~\cite{2017PhRvL.119a1101O,MandelOShaughnessy:2010,2013PhRvD..87j4028G,Stevenson:2015bqa,Farr:2017uvj,Stevenson:2017dlk,Gerosa:2017kvu,Fishbach:2017dwv,Talbot:2017yur}),
which also requires an understanding of the spin selection function~\cite{PhysRevD.82.104006, 2015ApJ...806..263D,Ng:2018neg}. 
The work presented here is complementary to these studies, as it aims to
incorporate an astrophysical distribution model as part of a mixture with
a noise component. As the observed population is limited by the sensitivity of Advanced
ground-based detectors, the population of candidate sources at the greatest
distances (lowest detection significance) will be contaminated with background
events.  We expect our method, using information from such sources, to improve both 
the precision and accuracy of merger rate and population parameter estimates; though 
as we will see, the degree of improvement depends on how easily the foreground
and background populations can be separated by existing analyses.

We start by defining our notation and deriving the general form of the model in
section~\ref{sec:notation_and_method}. Section~\ref{sec:toy_model} describes
its application to a toy model of mass distribution inference in the presence
of noise, and shows its application to a range of simple analytic population models.
In section~\ref{sec:er4} we consider a more realistic simulated data 
set derived from an engineering run prior to the start of Advanced LIGO 
observations in 2015.  We conclude in section~\ref{sec:conclusion}.

\section{Derivation of the generic model}
\label{sec:notation_and_method}

We consider a mixture of two populations, the astrophysical 
`foreground' and terrestrial noise `background': quantities defined analogously for
both populations will be distinguished by the subscripts $\mathrm{F}$ or $\mathrm{B}$ 
respectively.  Quantities without subscript will then refer to the total population 
which is the union of foreground and background. 

The model is also hierarchical: each event, if assumed astrophysical, has a set of
intrinsic properties such as component masses and spins, which we collectively
denote $\gamma$.
The distribution of these properties over each population 
is assumed to have a form described by a set of hyper-parameters.  We do not have access 
to the `true' values of properties for each event, only to a set of samples from a 
(typically Bayesian) estimate based on data around the event.  These samples are
derived under the assumption that the event is astrophysical, thus events that are 
in fact background will also be assigned parameter estimates.\footnote{We do not, 
of course, know with certainty that any given event is background.}

We then define the core quantities used in the following derivation as 
\begin{itemize}
    \item \snr, \vsnr \;: ranking statistic for one, resp.\ for all events in a given data set.
	\item \nobs, \nobsfg, \nobsbg \;: observed number of events above a threshold $\snr > \threshold$
	\item \nexp, \nexpfg, \nexpbg \;: expected number of events with $\snr > \threshold$, when modelling these as a Poisson process
    \item \popfg, \popbg \;: hyper-parameters which describe the population distributions
    \item \class, \vclass \;: flag indicating whether any given event, resp.\ all events, belong(s) to the astrophysical ($\eta = F$) or to the noise population ($\eta = B$)
    \item \PE, \vPE \;: vector of samples representing the parameter estimates (masses, spins, etc.) of one, resp.\ all events, under the assumption that events are astrophysical.
\end{itemize}

We wish to infer the joint posterior probability distribution of rates and population parameters for the two populations, given some events for which $\vsnr$ and $\vPE$ have been determined by the search and parameter estimation stages of data analysis:
\begin{align}
    p(\nexpfg, \nexpbg, \popfg, \popbg | \vsnr, \vPE, \nobs). \label{eq:pos}
\end{align}

Using Bayes' theorem we can express the posterior distribution \eqref{eq:pos} in terms of prior and likelihood functions,
\begin{align}
    p(\nexpfg, \nexpbg, \popfg, \popbg | \vsnr, \vPE, \nobs) 
      &= \frac{p(\vsnr, \vPE,  \nobs | \nexpfg, \nexpbg, \popfg, \popbg) p(\nexpfg, \nexpbg, \popfg, \popbg )} {p(\vsnr, \vPE , \nobs)}
\end{align}
We drop the normalisation constant $p(\vsnr, \vPE , \nobs)$ and factor out the likelihood for $\nobs$ as being independent of the population shape parameters,
\begin{align}
    p(\vsnr, \vPE,  \nobs | \nexpfg, \nexpbg, \popfg, \popbg) &= p(\nobs | \nexpfg, \nexpbg)p(\vsnr, \vPE | \nexpfg, \nexpbg, \popfg, \popbg) \nonumber \\
    &= \frac{\nexp^\nobs e^{-\nexp}}{\nobs!}  p(\vsnr, \vPE | \nexpfg, \nexpbg, \popfg, \popbg),
\end{align}
where we use a Poisson likelihood for $\nobs$ with a total expected number of events $\nexp = \nexpfg + \nexpbg$.
The second term, $p(\vsnr, \vPE | \nexpfg, \nexpbg, \popfg, \popbg)$, is the likelihood for the observed SNRs and parameter estimates, for the mixture model. We assume each event is conditionally independent given the population parameters, and so the joint likelihood is just the product of the likelihood for each one,
\begin{equation}
    p(\vsnr, \vPE | \nexpfg, \nexpbg, \popfg, \popbg)=\prod_i p(\snr, \PE | \nexpfg, \nexpbg, \popfg, \popbg).
\end{equation}
Now, we can split each of these into terms for the astrophysical and noise sub-models by introducing an indicator variable $\class\in(F,B)$, whose probability will depend on the rate parameters $\nexpfg$ and $\nexpbg$,
\begin{align}
    p(\snr, \PE | \nexpfg, \nexpbg, \popfg, \popbg) &= p(\snr, \PE | \popfg, \isfg)p(\isfg|\nexpfg,\nexpbg) \nonumber \\
        &\qquad + p(\snr, \PE | \popbg, \isbg)p(\isbg|\nexpfg,\nexpbg) \nonumber \\
    &= p(\snr, \PE | \popfg, \isfg)\frac{\nexpfg}{\nexp} + p(\snr, \PE | \popbg, \isbg)\frac{\nexpbg}{\nexp},
\end{align}
where the probability of each class is just the expected fraction of the total number. Since this is a sum of probability densities, special care must be taken to ensure all terms are properly normalised, such that
\begin{align}
	\int^\infty_{\threshold} \dd{\rho} \int \dd{\gamma} p(\rho, \gamma | \theta_\eta, \class) = 1,
\end{align}
for $\eta = \mathrm{F}$ and $\eta = \mathrm{B}$, where $\threshold$ is a minimum SNR value for which 
events are considered, either as a result of the event generation method or as a choice to limit 
computational costs.  
Neglecting this normalization would introduce an artificial preference for one component over the other. 
An extension to further sub-populations is simply achieved by including additional classes 
with their own rate and shape parameters. 

Recombining the pieces, we can write the desired posterior in Eq.~\eqref{eq:pos} as
\begin{multline}
    p(\nexpfg, \nexpbg, \popfg, \popbg | \vsnr, \vPE, \nobs) \propto \\
		p(\nexpfg, \nexpbg, \popfg, \popbg) e^{-\nexp} \prod_i \big[\, p(\snr, \PE | \popfg, \isfg)\nexpfg + p(\snr, \PE | \popbg, \isbg)\nexpbg \,\big]. 
\label{eq:generic_result}
\end{multline}
This expression is similar to Eq.(21) from \citet{FGMC} with an explicitly added dependence on source 
parameter estimates. 
This implies that our formalism reduces to the \citet{FGMC} result as used by LVC to estimate binary black 
hole merger rates~\cite{O1BBH,GW150914:rates}, if the event distribution over mass or similar parameters is 
not free to vary. 

The dependence on event parameters arises through the use of samples $\PE$, $i=1\ldots n$, drawn from the likelihood function of the data $d$ for a given point in parameter space $p(d|\gamma)$.  
These allow us evaluate the population likelihood function via marginalisation over the unknown true parameters, 
using the $n$ samples to perform a Monte Carlo integral as in~\cite{Mandel:2009pc},
\begin{align}\label{eq:PEsamples}
p(d|\popfg)&=\int p(d|\gamma)p(\gamma|\popfg) \dd{\gamma} \nonumber \\
&= \left< p(\PE|\popfg) \right>_{p(d|\PE)} \nonumber \\
&= n^{-1} \sum_j p(\gamma_j|\popfg). 
\end{align}
Samples from the likelihood therefore serve as a useful intermediate representation of the
raw interferometer data $d$.

To obtain a quantity directly relevant for an astrophysical interpretation, the expected number can be 
transformed into the local merger rate $\rate$ using the observing time $T$ and the sensitive volume $V(\gamma)$:
\begin{align}
    \rate = \frac{\nexpfg}{T \int \dd{\gamma} V(\gamma) p(\gamma|\popfg)},\label{eq:rate}
\end{align}
where the integral marginalises over the space of source parameters $\gamma$. In practice $V(\gamma)$ is estimated for a particular data-set by a Monte Carlo campaign, adding (`injecting') 
a large number of simulated signals to the data and counting the resulting number of events above threshold.

An additional quantity which is not directly used in the derivation of our model, but is important
for an astrophysical interpretation, is the probability of any given event originating from to the 
astrophysical foreground $p_\mathrm{astro}$:
\begin{gather}
	p(\isfg, \nexpfg, \nexpbg, \popfg, \popbg | \snr, \PE) = \frac{p(\snr, \PE | \popfg, \isfg) \nexpfg \times p(\nexpfg, \nexpbg, \popfg, \popbg)}{p(\snr, \PE | \popfg, \isfg) \nexpfg + p(\snr, \PE | \popbg, \isbg) \nexpbg}, \\
	p_{\mathrm{astro},i} = \int_0^\infty \dd{\nexpfg} \int_0^\infty \dd{\nexpbg} \int \dd{\popfg} \int \dd{\popbg} p(\isfg, \nexpfg, \nexpbg, \popfg, \popbg | \snr, \PE),
\end{gather}
where we marginalised over the population parameters.

\section{Toy model}
\label{sec:toy_model}
We construct a simple toy model of the universe to test our inference framework in various ways. 
The toy model allows us to generate a large number of realisations from the same underlying parameters, 
and to be certain that we use the correct model when analysing these realisations. 
For simplicity we consider a static, flat, and finite universe. 
Events are characterised completely by the distance $r$ to the source and a single mass parameter $m$,
which takes the place of the $\gamma$ used in the previously derived expressions. 
This mass parameter can be thought of as similar to the chirp mass. 
Additional effects such as inclination, spins, mass ratio, or antenna patterns are ignored. 

For our detection statistic $\rho$ we simply use (a simplified proxy for) the SNR, 
whose expected value $\truesnr$ is determined by $r$ and $m$ as
\begin{equation}
	\truesnr = K \frac{m}{r}.
\label{eq:snr_relation}
\end{equation}
where $K$ is an arbitrary constant which quantifies the detector sensitivity. 
We also model the uncertainty in the estimation of the mass parameter as
\begin{equation}
	\sigma_\mathrm{PE} \propto \frac{m}{\rho}
\label{eq:width_relation}
\end{equation}
which is a simplification of the relation given in~\cite{Cutler:1994ys}.

To apply the generic form found in Eq.~\eqref{eq:generic_result} to a specific problem, 
we need to evaluate the terms $p(\snr, \vec{m}_i | \popfg, \isfg)$ and $p(\snr, \vec{m}_i | \popbg, \isbg)$. 
This involves finding a functional form for the selection effects. 
For sources distributed uniformly in our static universe we can derive the needed expression directly 
by manipulating the joint distribution of masses and observed (detection) SNRs $\obssnr$.
We use the SNR relation defined above in Eq.~\eqref{eq:snr_relation}, which defines a mass dependent 
lower cut-off $\cutoff(m) = K\ m\ r_U^{-1}$ to the \truesnr\ possible in our toy universe.
Additionally we use the fact that the euclidean distances $r = |\vec{x}|$ of sources distributed uniformly 
in volume follow a $r^2$-distribution.  Therefore
\begin{align}
	p(\obssnr, m) &= \int_{V_U} \dd{\vec{x}} p(\obssnr, m | \vec{x}) p(\vec{x}) \nonumber \\
	&= \frac{1}{V_U} \int_0^{r_U} \dd{r} 4 \pi r^2 p(\obssnr, m | r) \nonumber \\
	&= \frac{1}{V_U} \int_0^{r_U} \dd{r} \int_{\cutoff(m)}^{\infty} \dd{\truesnr} 4 \pi r^2 p(\obssnr | \truesnr) p(\truesnr | m, r) p(m) \nonumber \\
	&= \frac{4 \pi p(m)}{V_U} \int_{\cutoff(m)}^{\infty} \dd{\hat{\rho}} \int_{\cutoff(m)}^{\infty} \dd{\truesnr} p(\obssnr | \truesnr)  \frac{(Km)^3}{\hat{\rho}^4} \delta(\hat{\rho} - \truesnr) \nonumber \\
	&= \frac{4 \pi K^3 m^3 p(m)}{V_U} \int_{\cutoff(m)}^{\infty} \dd{\truesnr} \rho_\mathrm{true}^{-4} \: p(\obssnr | \truesnr) \nonumber \\
	&\propto m^3 p(m) \int_{\cutoff(m)}^{\infty} \dd{\truesnr} \rho_\mathrm{true}^{-4} \: p(\obssnr | \truesnr),
\label{eq:component_with_se}
\end{align}
where $r_U$ and $V_U$ are the radius and volume of our universe, respectively, and 
$\cutoff(m) = \frac{K m}{r_U}$ is the lower SNR cut-off defined by a source of mass $m$ being placed at the 
maximum allowed distance $r_U$.  Thus, the SNR distribution for astrophysical sources is $p(\rho) \propto 
\rho^{-4}$, and we expect the observed mass distribution to be biased by a factor of $m^3$.  The mass and 
SNR components of Eq.~\eqref{eq:component_with_se} are generally connected via the mass dependent SNR cut-off. 
The term $p(\obssnr | \truesnr)$ accounts for the shift in search SNR relative to the expected value due to 
detector noise: 
\begin{equation}
	p(\obssnr | \truesnr) = \chi_\mathrm{NC}(\obssnr; \lambda=\truesnr, k=2), 
\label{eq:obs_snr}
\end{equation}
where $\chi_\mathrm{NC}$ is the non-central chi distribution with a non-centrality of $\lambda = \truesnr$ 
and $k=2$ degrees of freedom.

For real binary merger events we can pursue an analogous derivation, though the resulting relation differs as 
the SNR is a more complex function of event parameters than Eq.~\eqref{eq:snr_relation}. Additional 
complications arise if the detector is sensitive to events at cosmological distances, causing the observed 
masses to be redshifted by a distance-dependent amount.

The search and parameter estimation analyses that produce our events can only cover a finite range of masses, 
of which we denote the limits as $m_\mathrm{min}$, $m_\mathrm{max}$.  We will assume that all astrophysical 
foreground events have masses lying within these limits; in practice one should take sufficiently wide limits 
that the density of foreground events at these limits becomes vanishingly small.

For the mass distribution of the astrophysical foreground we consider two types of population distribution: 
a truncated power law
\begin{equation}
	p(m | \popfg) \equiv p(m | \alpha, m_\mathrm{low}, m_\mathrm{high}) \propto \begin{cases} m^\alpha 
	& \mathrm{if} \; m_\mathrm{low} < m < m_\mathrm{high} \\ 0 & \mathrm{else} \end{cases} \\
\label{eq:powerlaw_distribution}
\end{equation}
with three free parameters, the slope $\alpha$, lower mass cut-off $m_\mathrm{low}$, and high mass 
cut-off $m_\mathrm{high}$.  The two mass cut-offs are constrained by the 
mass range considered as $m_\mathrm{min} \le m_\mathrm{low} < m_\mathrm{high} \le m_\mathrm{max}$.  
The second population is a Gaussian
\begin{equation}
	p(m | \popfg) = p(m | \mu, \sigma) \propto \begin{cases} \mathcal{N}(m; \mu, \sigma) & \mathrm{if} \; 
 m_\mathrm{min} < m < m_\mathrm{max} \\ 0 & \mathrm{else} \end{cases} \\
\label{eq:gaussian_distribution}
\end{equation}
with two free parameters, the mean $\mu$ and standard deviation $\sigma$.  Strictly this distribution is a 
truncated Gaussian, however in practice we consider parameter ranges such that $p(m | \popfg) \ll 1 $ at 
the boundaries. 
In contrast to the explicit differentiation between the true and observed SNR,
Bayesian parameter estimation provides us with samples from the probability
distribution of the true mass which are used directly as in Eq.~\eqref{eq:PEsamples},
which eliminates the need to introduce a variable representing an observed mass.

In the background case, there are no selection effects, and we assume the noise characteristics are such 
that there is no correlation between the mass distribution and the SNR distribution. As a result, 
$p(\snr, \vec{m}_i | \popbg, \isbg)$ decomposes as simply
\begin{equation}
	p(\snr, \vec{m}_i | \popbg, \isbg) = p(\snr | \popbg, \isbg) p(\vec{m}_i | \popbg, \isbg).
\end{equation}
Note that in realistic data the SNR distribution of background events may be strongly dependent
on the mass (and other template parameters)~\cite{GW150914:CBCsearch} so this decomposition is not 
necessarily valid.

The expected rate and distribution of background events caused by instrumental noise can, in practice, 
be measured to high precision using techniques such as time-shifted 
analyses~\cite{Capano:2016uif,Usman:2015kfa} (see also~\cite{Cannon:2012zt}). 
For our artificial universe we have the freedom to choose the SNR and mass distributions,
though this choice was informed by observed distributions in real data.
We choose a power law with slope $-12$ in SNR; the mass posteriors are of constant width with their 
central values distributed uniformly between $m_\mathrm{min}$ and $m_\mathrm{max}$,
\begin{align}
    p(\rho | \eta \! = \! \mathrm{B}) &\propto \rho^{-12}, \\
    p(m | \eta \! = \! \mathrm{B}) &\propto 1. 
\label{eq:bg_distributions}
\end{align}
More realistic choices would include the effect of template bank density~\cite{Dent:2013cva}
and transient noise glitches~\cite{Nitz:2017svb,Nitz:2017lco} on the distribution of noise triggers over 
mass space. 
Note that our inference of the foreground mass distribution is expected to become more precise 
the more distinct the foreground and background are, especially in SNR.
Here, both SNR distributions are falling power laws, however background drops off much more rapidly than foreground. 

Finally we combine the mass distribution with Eqs.~\eqref{eq:component_with_se}-\eqref{eq:obs_snr}. 
Using \eqref{eq:powerlaw_distribution} for the truncated power law we obtain:
\begin{align}
	&p(\rho, m | \alpha, m_\mathrm{low}, m_\mathrm{high}, \eta \! = \! \mathrm{F}) \nonumber \\
	&\propto \int_{\cutoff(m)}^{\infty} \dd{\truesnr} \rho_\mathrm{true}^{-4} \: \chi_\mathrm{NC}(\rho; \lambda\!=\!\truesnr, k\!=\!2) \times \begin{cases} m^{\alpha+3} & \mathrm{if} \; m_\mathrm{low} < m < m_\mathrm{high} \\ 0 & \mathrm{else} \end{cases}
\end{align}
Using \eqref{eq:gaussian_distribution} for the Gaussian:
\begin{align}
	&p(\rho, m | \mu, \sigma, \eta \! = \! \mathrm{F}) \nonumber \\
	&\propto \int_{\cutoff(m)}^{\infty} \dd{\truesnr} \rho_\mathrm{true}^{-4} \: \chi_\mathrm{NC}(\rho; \lambda\!=\!\truesnr, k\!=\!2) \times \begin{cases} m^3\mathcal{N}(m; \mu, \sigma) & \mathrm{if} \; m_\mathrm{min} < m < m_\mathrm{max} \\ 0 & \mathrm{else} \end{cases}
\end{align}
The background model does not involve selection effects and yields
\begin{equation}
	p(\rho, m | \eta \! = \! \mathrm{B}) \propto \rho^{-12}. 
\end{equation}
In general normalising these expressions requires an integral over $\rho$ and $m$ which can be difficult 
or computationally expensive.  In our model this simplifies somewhat as the integrand 
$\rho_\mathrm{true}^{-4} \: \chi_\mathrm{NC}(\rho; \lambda\!=\!\truesnr, k\!=\!2)$ happens to assume values 
very close to zero for the \cutoff\ values of $[0.25, 4]$ allowed by our prior mass range of $[5, 80]$, 
therefore we are able to approximate $\cutoff = 1$.

To generate the artificial datasets we draw a total number of foreground and background events 
from a Poisson distribution around the true values determined by the intrinsic rate. Each of 
those events corresponds necessarily to a local maximum of signal likelihood over time, mass, 
and, in general, other parameters - we generally approximate this local maximum as a multivariate 
Gaussian distribution. For each foreground event we then draw the true mass and distance, from 
which we can uniquely determine the intrinsic SNR. We then simulate the impact of noise on the 
measurement of both SNR and mass by drawing a value from a non-central chi distribution around 
the true SNR to obtain the observed SNR, and drawing the maximum likelihood value from a normal 
distribution around the true mass with a width as determined by Eq.~\eqref{eq:width_relation}. 
We use a uniform in mass prior, therefore the posterior samples for the mass estimate are drawn 
from a Gaussian with the same width around the maximum likelihood value. 
For background events the observed SNR is drawn directly from a 
power-law as the background SNR distribution is determined empirically from the observed SNR 
values, while the posterior samples for the mass are drawn from a constant-width Gaussian 
around a central value drawn from a uniform distribution between $m_\mathrm{min}$ and 
$m_\mathrm{max}$ for each event.

Our method does not require us to make strong assumptions about the shape or width of the mass likelihood, 
however it is important to generate these artificial results carefully as negligence can have unexpected consequences. 
In practice, we have found the scaling and width of the posteriors to have little effect on our results 
when the posteriors are of smaller scale than the population.

\section{Toy Model Results}

We applied our method to a large number of realisations for each choice of foreground distribution, 
though the figures in the following section only show results for a single realisation. 
The results across realisations will be given in text only.  
The mass limits chosen for all toy model results\footnote{Since we do not claim a specific link to 
astrophysics in the toy model, the mass units are arbitrary.} were $m_\mathrm{min} = 5$, $m_\mathrm{max} = 80$.
The total expected number of events above an SNR of $8$, the lowest threshold considered, is $1600$, with 
$95\%$ contamination due to background events. 
The chosen slope of $-12$ for the background SNR distribution is less steep than in typical LIGO-Virgo
analyses for stellar-mass compact binary mergers; our choice exaggerates the transition region in which the 
chances of an event belonging to either foreground or background are comparable. 

To simulate the limitations due to the computational costs of the analysis we impose a SNR threshold on 
events, assessing its influence on our inferences by varying its value between $8$ and $15$. 
These SNR values would typically be considered as sub-threshold, since an SNR of $\approx 13.7$ is required 
for an event to have a $p_\mathrm{astro}$ value of $50\%$, and to reach $p_\mathrm{astro} = 90\%$ an SNR of 
$\approx 18$ is needed.
Lastly, the free parameters in Eq.~\eqref{eq:snr_relation} and Eq.~\eqref{eq:width_relation} are chosen
such that an event of true mass $m = 30$ at a notional distance of $400$\,Mpc has a mass posterior with 
width $\sigma_\mathrm{PE} \approx 1$, and has an SNR of $\rho \approx 50$.  
The width of the mass posteriors of background events is set to the constant value of $3.2$, typical of 
foreground events at the lowest SNR considered.

The priors chosen are flat in all hyper-parameters, with two exceptions: the 
width of Gaussian populations, where the prior was flat-in-log, and the expected number of astrophysical 
foreground events, where we used a Jeffreys prior:
\begin{align}
	\mathrm{prior}(\sigma) &\propto \frac{1}{\sigma}, \\
	\mathrm{prior}(\nexpfg) &\propto \frac{1}{\sqrt{\nexpfg}}.
\end{align}
Parameter estimation was performed using the \texttt{emcee}~\cite{emcee} implementation of an Affine Invariant 
Markov chain Monte Carlo Ensemble sampler~\cite{GoodmanWeare}.

\subsection{Power law distribution}
\label{sec:powerlaw}

The first population considered was the truncated power law, which was inspired by the 
idea that black hole masses may be distributed analogously to the initial mass function
of their progenitor stars.
We add parameters $m_\mathrm{min}$ and $m_\mathrm{max}$ to define the lower and upper limits
of the power law distribution. This is motivated by the desire to determine whether there
are gaps in the astrophysical black hole mass distribution: at the low end to compare
with the apparent lower limit of black hole mass in X-ray binaries~\cite{Farr:2010tu}, and at
the high end to determine the maximum mass above which a pair-instability supernova completely
disrupts the star~\cite{1967PhRvL..18..379B}.
In our simulation we chose the power law slope to be $-2.4$, and the cut-off values to be $12$ and $64$.
\begin{figure}
	\includegraphics[width=0.8\textwidth]{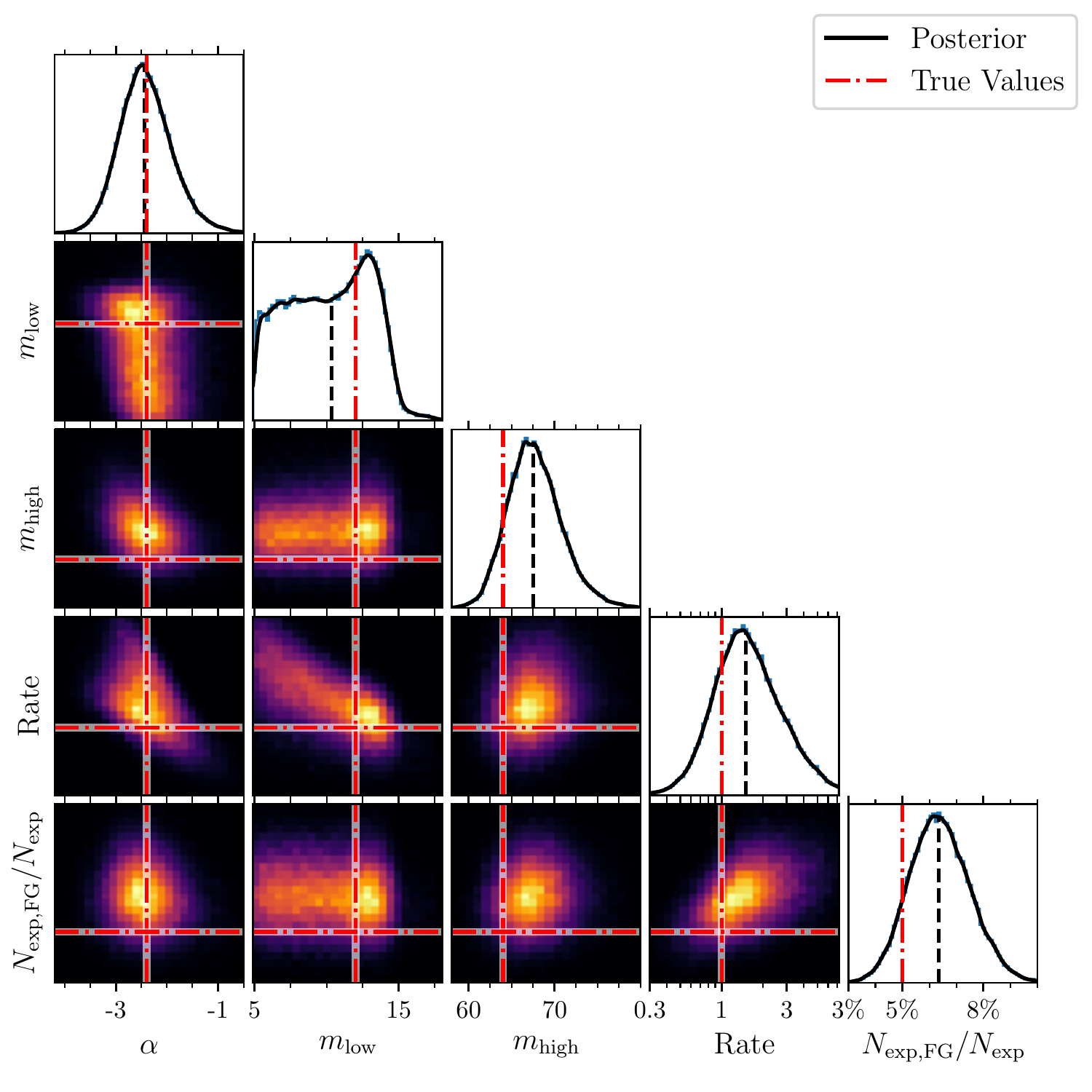}
	\centering
	\caption{Parameter estimates for a single realisation of the toy model. 
			The foreground population model is a truncated power law with 
			slope $\alpha$, and cut-offs $m_\mathrm{min}$ and $m_\mathrm{max}$.
			The expected number of events above the SNR threshold of 8 are 1620, 5\% of which are 
			expected to be foreground events.
			The black lines show the kernel density estimate of the posterior, the thin blue lines the corresponding histogram.
			The red dash-dotted line indicates the true value for the underlying population.
			}
	\label{fig:powerlaw_corner}
\end{figure}

Our primary results, the estimates of model parameters and their correlations, are shown in figure~\ref{fig:powerlaw_corner}. 
These results use a SNR threshold 8, the lowest value for which we run our analysis, as we would
expect this to yield the best possible parameter estimates.
Notable features are the large spread of possible merger rate densities (abbreviated as ``Rate'') 
and their correlation with the lower mass cut-off.  
This is a consequence of the fact that the power law slope is effectively increased by $3$ due to selection 
effects, thus detected events are described by a positive slope.
The detection bias towards high mass means that fewer events are available to define the lower cut-off value, 
and low mass events which are observed tend to have lower SNR values.  
As the total rate is still dominated by low mass (and low-amplitude) events,
 the large uncertainty of the low mass cut-off yields a high uncertainty on the rate. 
The estimated fraction of foreground events in the sample of observed events couples linearly to the merger 
rate density, but is less significant than the lower mass cut-off.
\begin{figure}
\centering
	\includegraphics[width=0.49\textwidth]{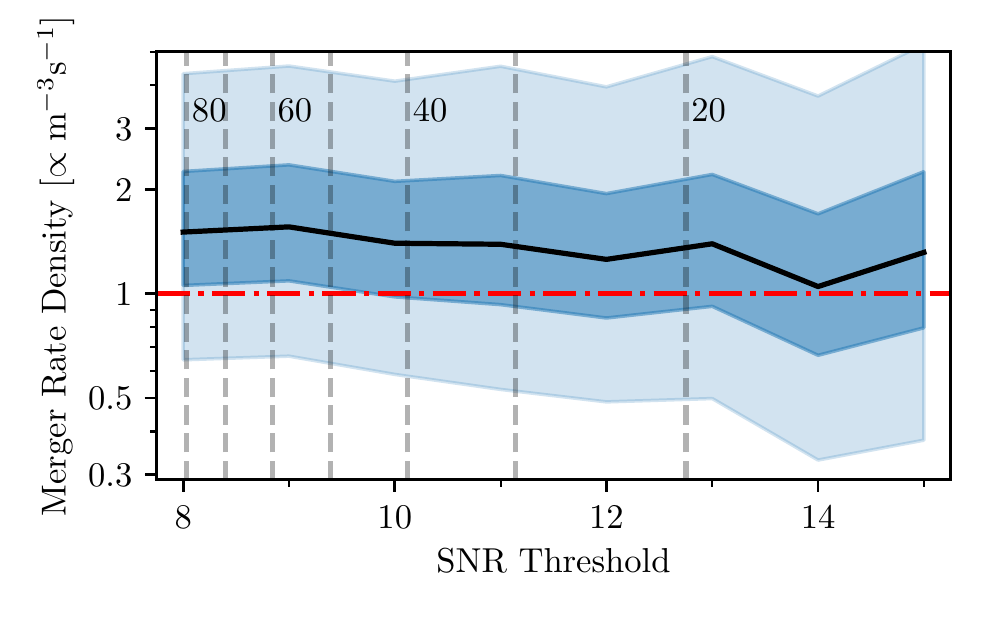}
	\includegraphics[width=0.49\textwidth]{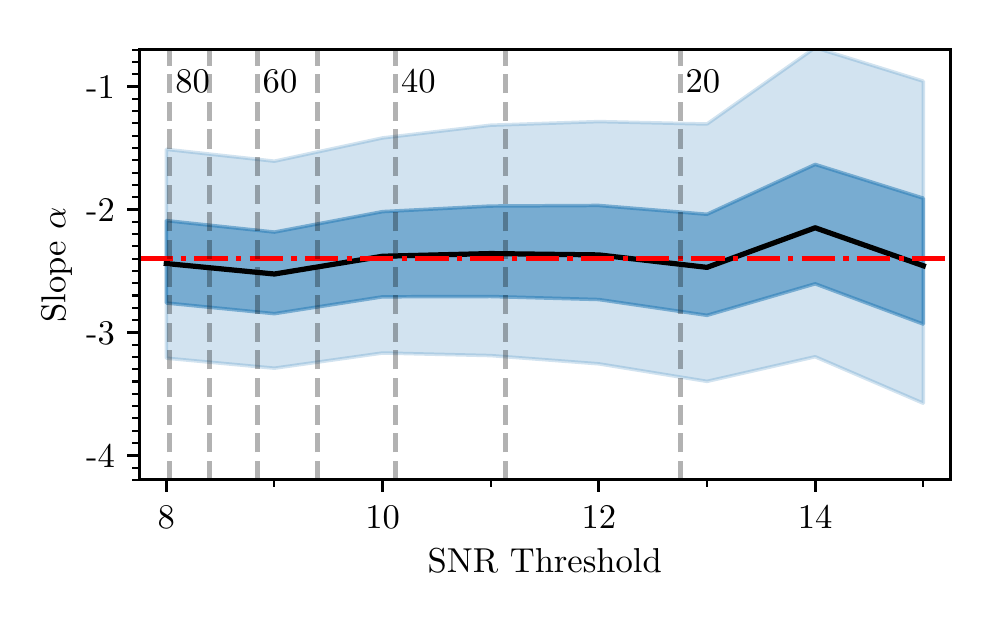} \\
	\includegraphics[width=0.49\textwidth]{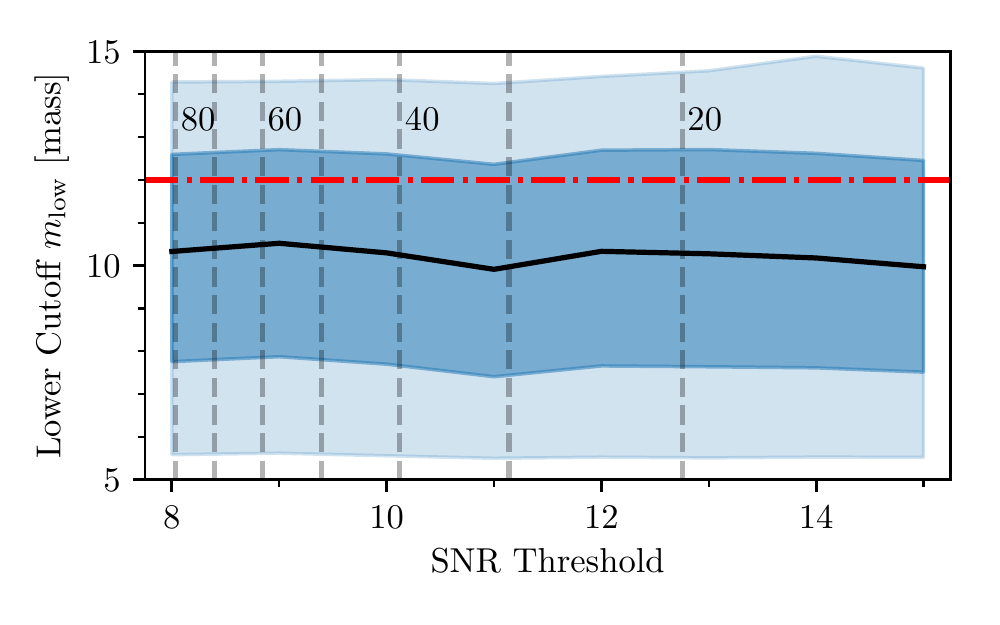}
	\includegraphics[width=0.49\textwidth]{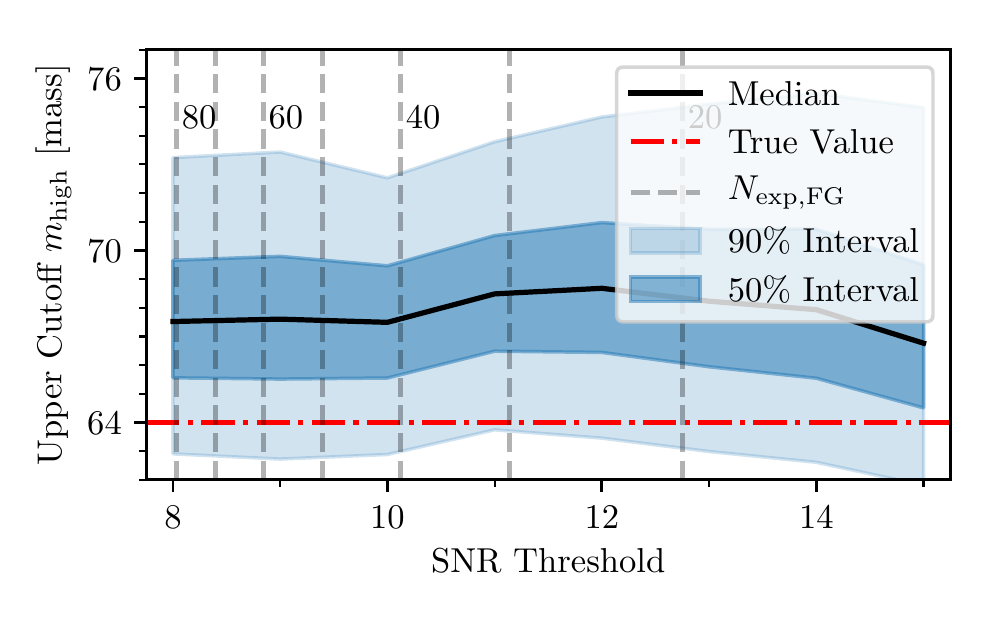}
	\caption{Confidence intervals for individual parameters of one realisation of the truncated power-law model, 
			as a function of SNR threshold. 
			The parameters shown are the inferred astrophysical merger rate (upper left) and power law slope (upper right), 
			as well as the low (lower left) and high (lower right) mass cut-offs.
			The red dash-dotted line indicates the true value for the underlying population.
			Dashed grey lines indicate the expected number of foreground events at the given SNR threshold.}
	\label{fig:powerlaw_bands}
\end{figure}

To assess our method in the light of its main goal of avoiding bias while lowering the SNR threshold, 
a single analysis result is insufficient.  Therefore, we analyse the same data with a range of different SNR 
thresholds to observe the change in the hyper-parameter estimates. Figure~\ref{fig:powerlaw_bands} shows 
the marginalised posteriors for the rate and power law slope as a function of SNR threshold. We can observe 
the posteriors growing slightly wider as the SNR threshold is increased and information from fewer events 
is considered.  The result from
one single realisation is, however, not necessarily representative of the general behaviour.
Combining the results from multiple realisation shows there is no noticeable bias regardless of the threshold 
chosen, and estimates improve slightly as the threshold is lowered. 
The width of the 90\% credible intervals decreases by factors of $\approx 1.5$ for the power-law slope, 
$\approx 1.1$ and $\approx 1.05$ for the lower and upper mass cut-offs respectively, 
and $\approx 1.2$ for the log of the inferred merger rate density.
\begin{figure}
	\includegraphics[width=0.65\textwidth]{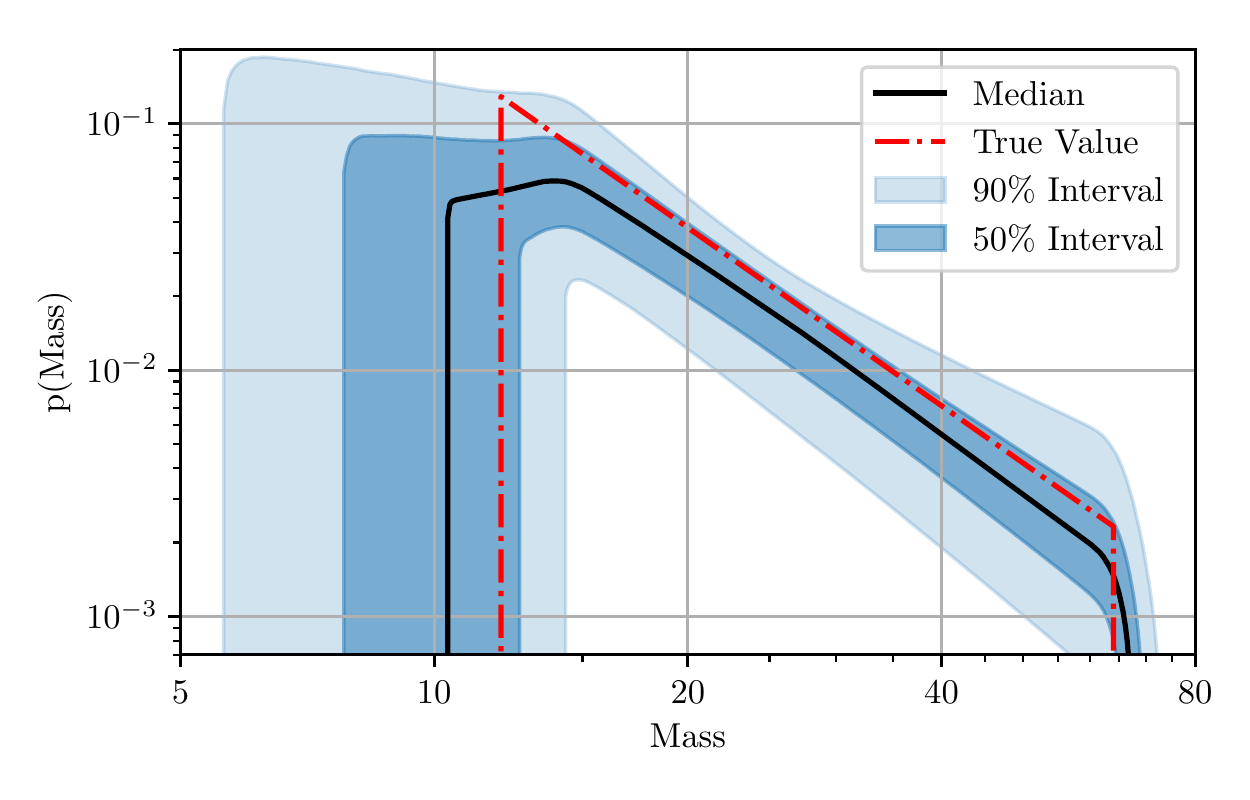}
	\centering
	\caption{The inferred mass distribution of the foreground population.
			The bands indicate the given percentiles in the probability density at any given mass 
			across all posterior samples.}
	\label{fig:powerlaw_inferred}
\end{figure}

Given the estimates of population parameters we can also compute an estimate of the underlying mass distribution, which we show in figure~\ref{fig:powerlaw_inferred}. Here we show the $50\%$ and $90\%$ confidence bands, as defined by computing the percentiles of $p(m | \theta)$ across all samples $\theta$ from the posterior for any given mass $m$. We observe that the true distribution is contained well within the credible interval and deviations are generally caused by an underestimated lower cut-off. In general there is a trade-off between expanding the bounds of the mass distribution to include additional events, and shrinking it to increase the PDF for highly significant events. The lower cut-off tends to have more freedom of movement as there are fewer high SNR events at low mass to constrain it.
\begin{figure}
	\includegraphics[width=0.7\textwidth]{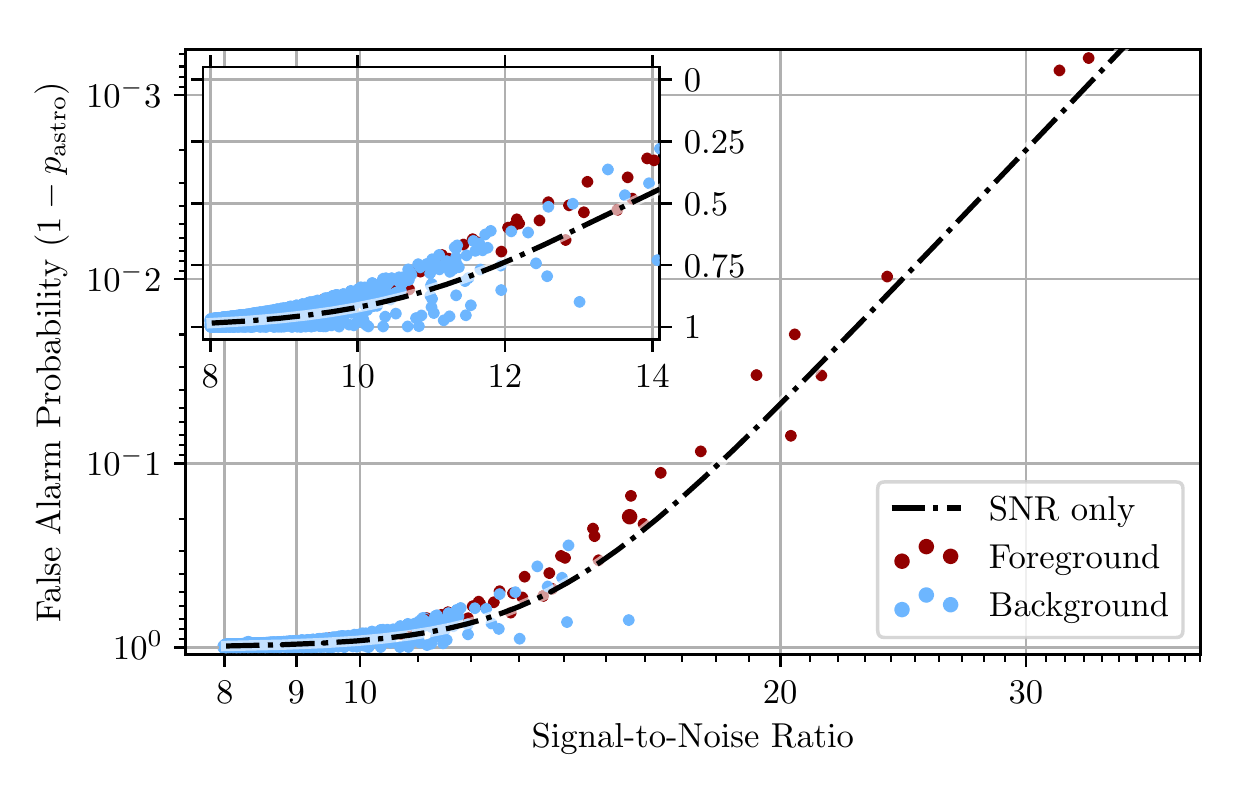}
	\centering
	\caption{The probability that an event is caused by detector noise rather than being of astrophysical origin, 
			$1 - p_{\mathrm{astro}}$, versus SNR.
			The foreground population model is a truncated power-law distribution.
			Blue and red dots represent foreground and background events respectively.
			The dash-dotted line shows the probability that would be inferred by a SNR based estimate~\cite{FGMC},
			assuming the relative number of expected foreground and background events is known perfectly.
			The inset focuses on the region with background events and emphasises events which are unlikely to be 
			astrophysical by using a linear scale.}
	\label{fig:powerlaw_fap}
\end{figure}

As a final result for this population, figure~\ref{fig:powerlaw_fap} shows the estimated probability 
of any given event to have an astrophysical origin $p_\mathrm{astro}$, and how it compares to a SNR-only 
estimate indicated by the black dash-dotted line. 
While this figure does not show quantitative results, we do observe that foreground events are largely 
located above the dash-dotted black line, indicating that our confidence in them being real has increased, while 
background events tend to be located below and are often on the $p_\mathrm{astro} = 0$ line when their 
masses are outside the hard cut-offs of the truncated power law population.
Thus we find, as expected, that the discrimination between signal and noise populations is improved with 
the incorporation of information about their mass distributions~\cite{Dent:2013cva}.
We determine that the percentage point difference between $p_\mathrm{astro}$ using our method and the SNR-only 
approach to be $\approx 3\%$, although this includes events with tiny absolute shifts due to them being 
very close to either $0\%$ or $100\%$ in the first place.

\subsection{Gaussian distribution}

The second core population considered is a simple Gaussian with a very small width.
This population was chosen to test the inference on hard-to-infer parameters and to see the 
effect a very distinctive distribution has on the discriminating power of our method. 
As such, we chose the width to be very narrow with a standard deviation of $1.6$ around a mean mass of $27$. 
This means that the width of the population is smaller than the typical width of the mass posterior of 
$\approx 2-3$ 
for each individual event, which we expect to be hard to estimate with a relatively small number of events. 
On the other hand, the discriminating power of using information from the mass estimates should be much 
greater than for a wide distribution like the truncated power law used in the previous section.
\begin{figure}
	\includegraphics[width=0.8\textwidth]{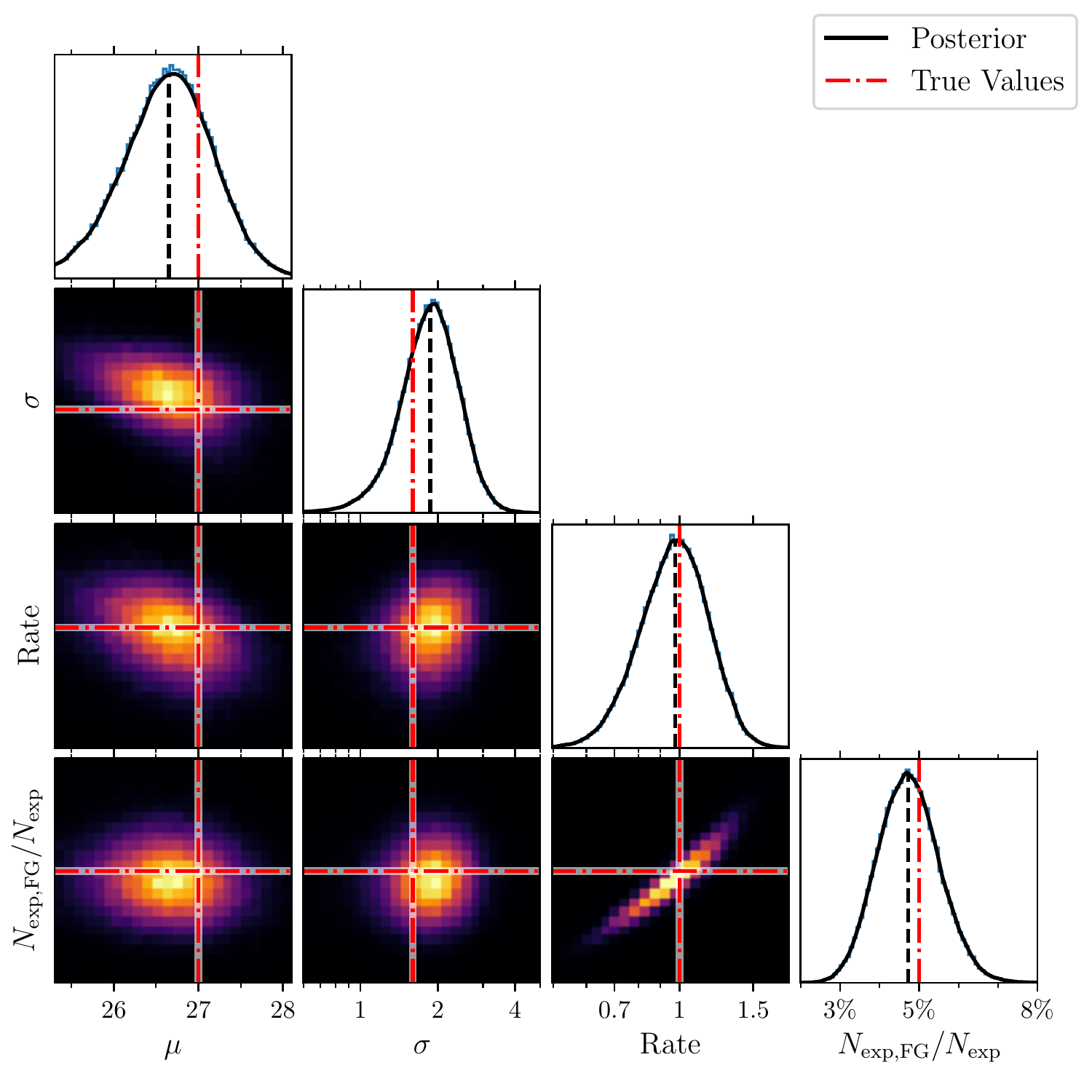}
	\centering
	\caption{Parameter estimates for a single realisation.
			The foreground population model is a Gaussian with two free parameters, the mean $\mu$ and width $\sigma$.
			The true expected numbers of events above a SNR threshold 8 is 1620, 
			$5\%$ of which are expected to be foreground events.
			The black lines show the kernel density estimate of the posterior, the thin blue lines the corresponding histogram.
			The red dash-dotted lines indicates the true values for the underlying population.}
	\label{fig:gaussian_corner}
\end{figure}

The parameter estimates for a single realisation are shown in figure~\ref{fig:gaussian_corner}. 
We observe that true width of the population $\sigma$ in contained comfortably within the inferred posterior, 
though the uncertainty is rather large. It is generally overestimated slightly.
Similarly the mean of the population is found well with an uncertainty comparable to the population width. 
The rate is constrained much better than in case of the truncated power law as this model lack the 
degeneracy between the rate and a poorly constrained population parameter.
The lack of a strong correlation between a population parameter and the merger rate density 
also highlights its linear relation to the estimated number of foreground events contained
within the sample.

When lowering the SNR threshold from 15 down to 8, the sizes of the $90\%$ confidence intervals 
of the population parameters and merger rate density decrease by median factors of $\approx 1.3-1.7$.
%
\begin{figure}
	\includegraphics[width=0.65\textwidth]{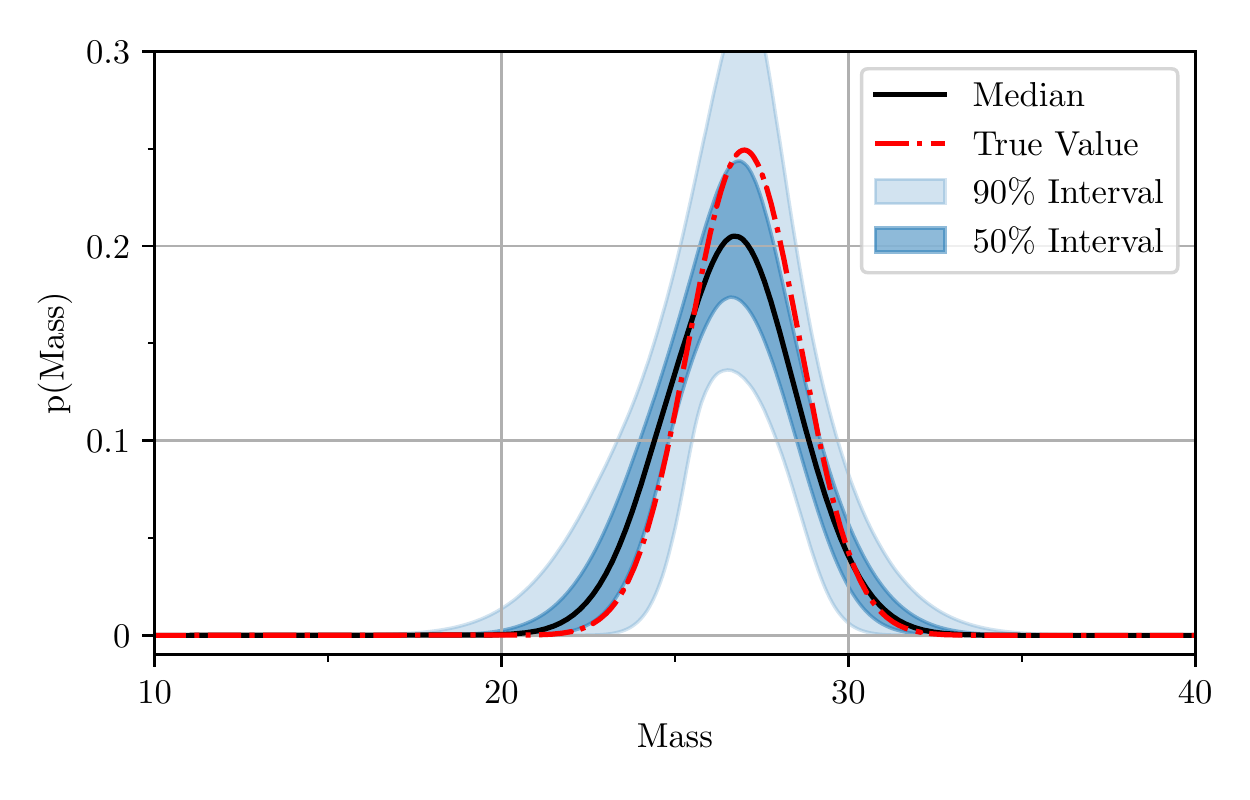}
	\centering
	\caption{The inferred mass distribution of the foreground population.
			The bands indicate the given percentiles in the probability density at any given 
			mass across all posterior samples.}
	\label{fig:gaussian_inferred}
\end{figure}
The true mass distribution is well within the confidence interval shown in figure~\ref{fig:gaussian_inferred}, though the true distribution is somewhat more narrow than inferred as seen previously in figure~\ref{fig:gaussian_corner}.
\begin{figure}
	\includegraphics[width=0.7\textwidth]{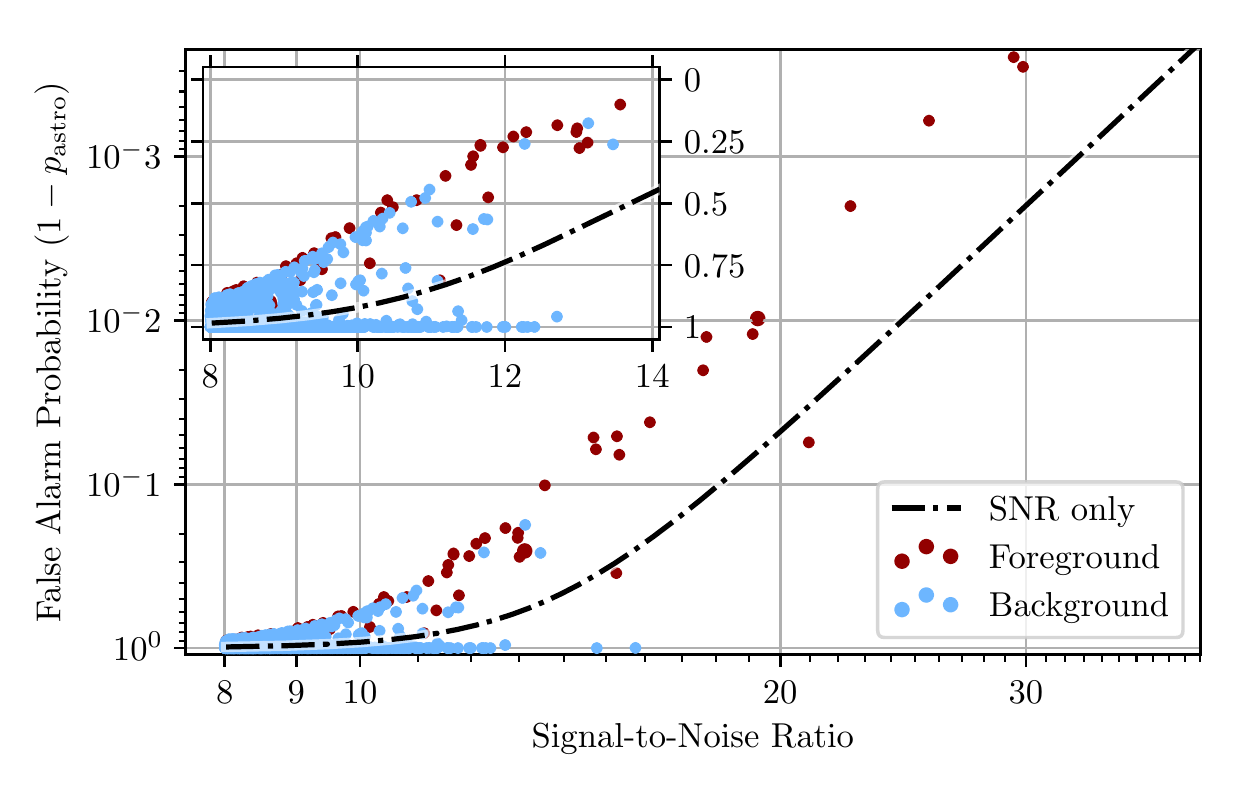}
	\centering
	\caption{The probability of an event being caused by detector noise rather than being of 
			astrophysical origin, $1 - p_{\mathrm{astro}}$, versus SNR. 
			The foreground population model is a narrow Gaussian distribution.
			Blue and red dots represent foreground and background events respectively.
			The dash-dotted line shows the probability that would be inferred by a SNR based 
			estimate~\cite{FGMC}, assuming the relative number of expected foreground and background 
			events is known perfectly.
			The inset focuses on the region with background events and emphasises events which are unlikely 
			to be astrophysical by using a linear scale.
			}
	\label{fig:gaussian_fap}
\end{figure}

The comparison of the estimated $p_\mathrm{astro}$ as shown in figure~\ref{fig:gaussian_fap} show 
how important the inclusion of masses is for this population. 
We can clearly identify the band of foreground events for which $1 - p_\mathrm{astro}$ is smaller 
by factors of a few up to 10 compared to the SNR based estimate. 
Only three out of $\approx 74$ foreground events lost any $p_\mathrm{astro}$. 
There are some background events which, too, have their $p_\mathrm{astro}$ increased, 
though most are demoted and often down to effectively $0$.

\subsection{Incorrect models - No background component}
\label{sec:no_noise}

Previous analyses of gravitational wave populations (such as the power law model used in 
\cite{GW170104}) use a high threshold to ensure a high probability that the events used are 
of astrophysical origin, in effect neglecting the possibility of background. Here we 
investigate the behaviour of our toy model with the background component disabled, corresponding
to such a scenario.
This shows the results one would obtain if simply fitting the foreground model to a contaminated dataset.
The underlying population is a truncated 
power law identical to the one used in the first set of results presented in section~\ref{sec:powerlaw}.
\begin{figure}
    \centering
	\includegraphics[width=0.49\textwidth]{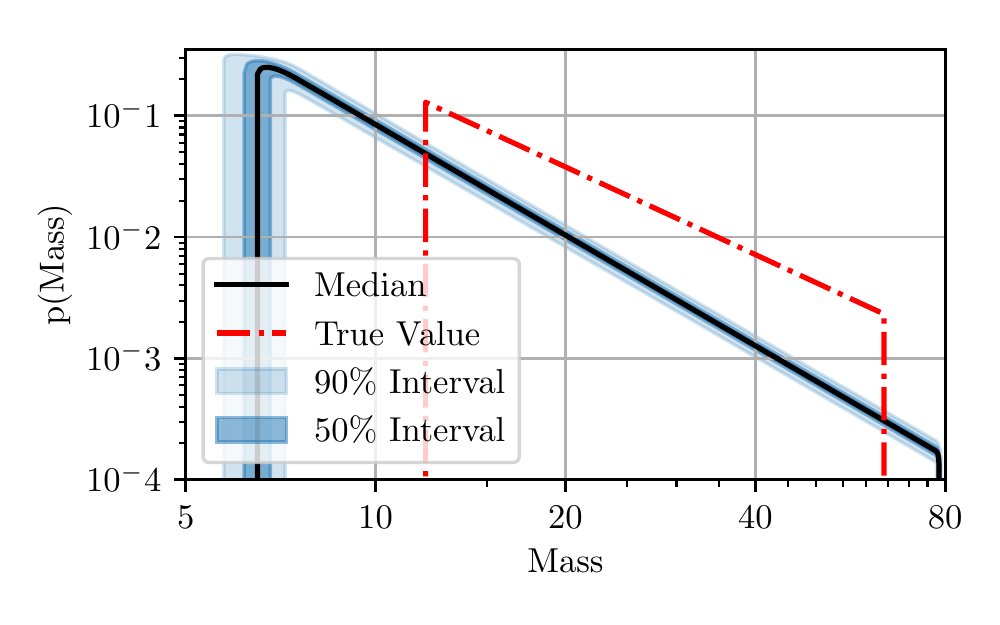}
	\includegraphics[width=0.49\textwidth]{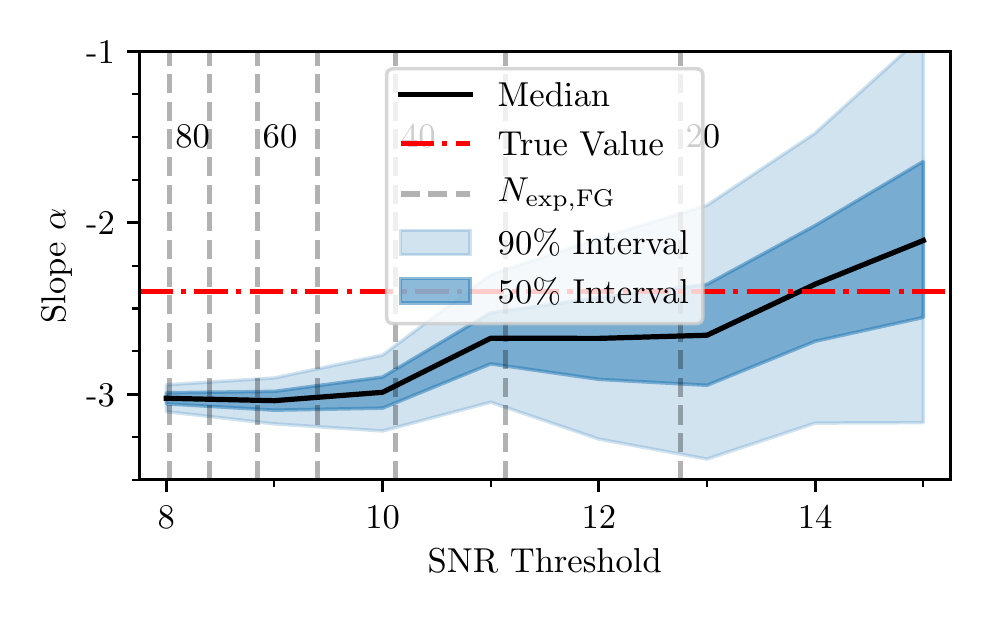}
	\caption{Results for the power law model when we neglect to account for contamination due to noise. 
			Left: Inferred mass distribution at the lowest SNR threshold of 8, the PDF percentiles are 
			calculated across the population posterior at any given mass.
			Right: Inferred power law slope as a function of SNR threshold. Vertical dashed lines 
			indicate the expected number of true events above a particular SNR.}
	\label{fig:no_noise_inferred}
\end{figure}

The results are shown in Fig.~\ref{fig:no_noise_inferred}, where we observe the inferred distribution to 
be very different from the true one when the lowest SNR threshold of $8$ is used and the dataset is 95\%
polluted (left panel).
The mass cut-offs are extended to the edges of the prior ranges to incorporate noise events at those values. 
The confidence interval includes the true value as long as the SNR threshold is sufficiently high 
since the number of background events is negligible, but trends towards $-3$ as the threshold is lowered. 
This is expected since the background dominates the low SNR region and has a power law slope of $0$ in mass. 
This slope corresponds to an actual slope of $-3$ when selection effects are considered.
In the right panel we see the effect on the estimation of the power law slope as the threshold is varied.
At a pollution fraction of 50\% (SNR 13.7) the statistical uncertainty of the slope becomes large enough that
the systematic bias is not visible.

\subsection{Incorrect models - Neglected selection effects}

The second kind of error we considered was the neglect to properly account for the mass dependence of 
selection effects.  In case of a power law distribution this is trivial, as it simply adds $+3$ to the 
inferred value of the slope.  Therefore we chose a Gaussian as the population, and we increased the 
width to $9$ to highlight the impact of selection effects on the inferred population.
\begin{figure}
	\includegraphics[width=0.65\textwidth]{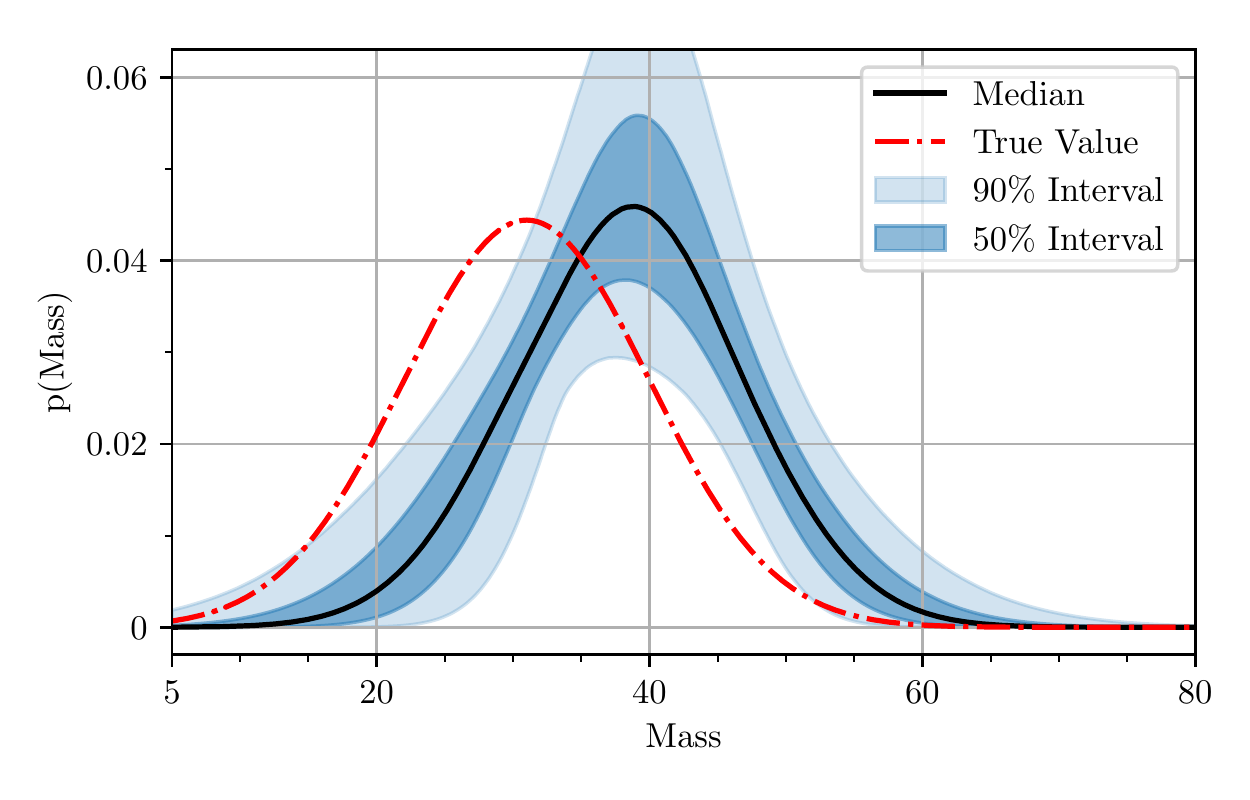}
	\centering
	\caption{The inferred mass distribution for the Gaussian model without compensating for the mass 
	dependence of selection effects, using the lowest SNR threshold of 8.  The PDF percentiles are 
	calculated across the population posterior at any given mass.}
	\label{fig:no_SE_inferred}
\end{figure}
Figure~\ref{fig:no_SE_inferred} shows that the selection effects effectively shift the distribution towards 
higher masses.
This is a general feature as the $m^3$ term strongly favours high mass events in the observed set of events.
Depending on the population the may also affect the width of the population, which happened to be 
a very minor effect in this case.  

Together with the previous section~\ref{sec:no_noise} this illustrates that population inference 
can be made impossible even when the model matches the underlying distribution. 
Accounting for the presence of noise and selection effects is essential for correct inference and 
to avoid bias when attempting to lower the SNR threshold.

\section{Advanced LIGO engineering run}\label{sec:er4}
After the successful tests using the simple toy model we also applied this method to more realistic data. 
For this we used simulated data from the fourth advanced LIGO engineering run (ER4).  
LIGO Hanford noise data were simulated as Gaussian noise using the LIGO design sensitivity noise 
spectrum~\cite{ObservingScenarios}, while LIGO Livingston data were derived from an instrumental channel
monitoring the power-stabilized laser, recoloured to the same average target spectrum. 
Simulated gravitational wave signals were then added (``injected'') to the recorded noise data streams.
The injected population of binary black hole mergers was chosen to be uniform in both component masses 
between limits of $5$ and $20~\Msun$, and uniform in volume with no cosmological effects included.  The 
{\tt EOBNRv2HM} approximant tuned to numerical relativity~\cite{Pan:2011gk} was used to simulate binary 
black hole mergers including non-dominant GW emission modes, for non-spinning binary components.  The 
intended astrophysical rate corresponding to the injected merger signals was $5\,\mathrm{Gpc}^{-3}
\mathrm{yr}^{-1}$.\footnote{Due 
to a software error the amplitude of injected signals was a factor 2 higher than intended, effectively
simulating a true merger rate of $40\,$Gpc$^{-3}$yr$^{-1}$; however in the results presented here, we 
rescale our rate estimates to compensate for this error.}

We searched the data for signals using the \texttt{pycbc}~\cite{Canton:2014ena,Usman:2015kfa,
pycbc-software,GW150914:CBCsearch} pipeline to cover binary mergers of non-spinning components with masses 
between $3$ and $50~\Msun$; this range also defined the prior for parameter estimation performed on each 
event using \texttt{LALinference}~\cite{lalinference}.
The search detection statistic for candidate events, $\rho_c$, is the quadrature 
sum of $\chi^2$-reweighted SNRs $\hat{\rho}_{H,L}$ over single-detector events
having consistent component masses and times of arrival between the two detectors~\cite{ihope,S6upperlimits}. 
The number of events we chose to analyse is limited by computational cost; we impose a threshold 
$\rho_c > 8$ leaving us with 100 events in $\approx 37$ days of LHO-LLO coincident observing time; 51 of 
these events correspond to known injected signals, with the remainder due to noise fluctuations.\footnote{In
reality we will not have access to an independent record listing all true signals!}

\begin{figure}
	\includegraphics[width=0.65\textwidth]{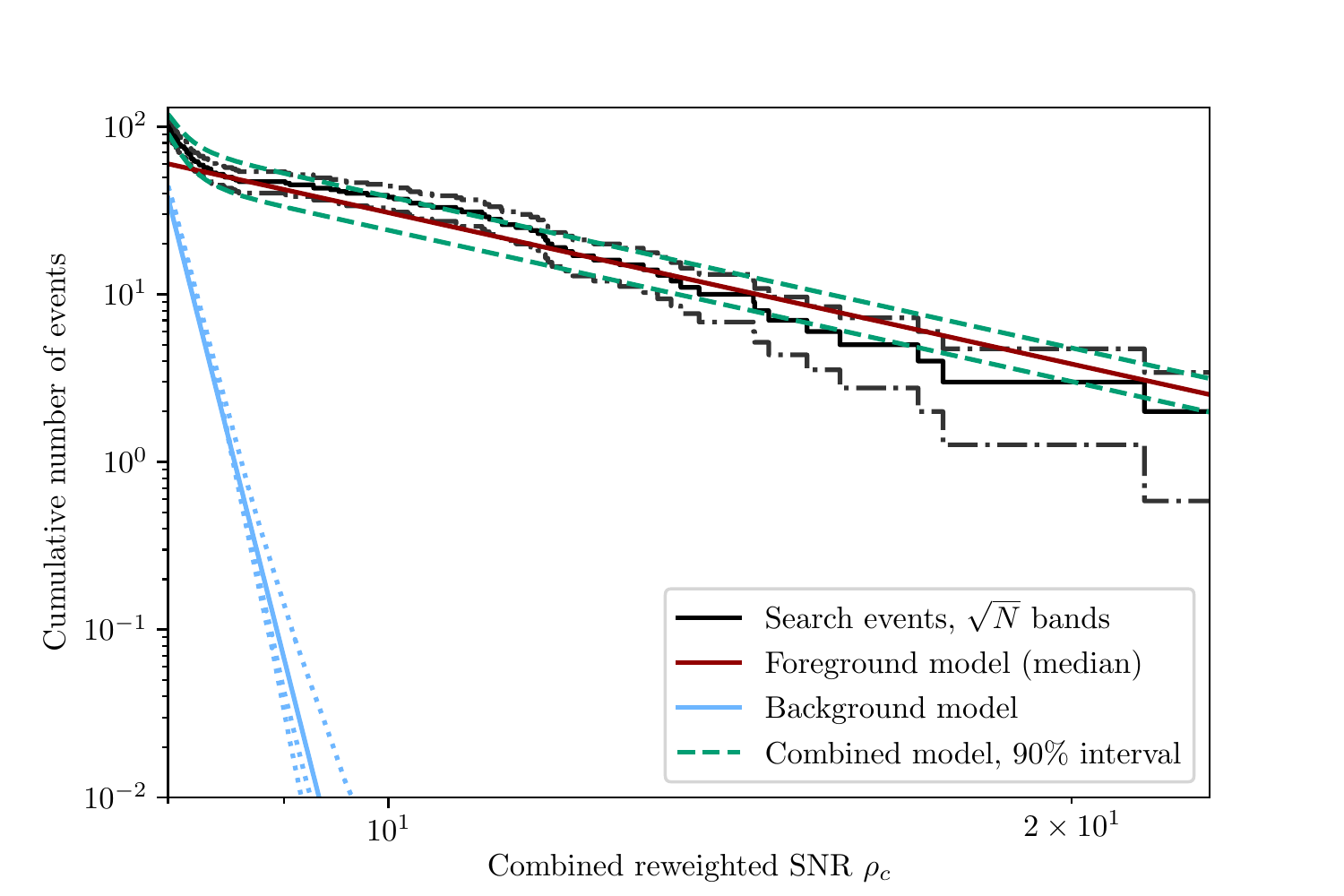}
	\centering
	\caption{Cumulative number of (simulated) events detected in LIGO ER4 engineering run data vs.\ 
	threshold search  
	detection statistic $\rho_c$.  Black steps indicate the search results, with $\pm\sqrt{N}$ bands 
	indicating expected counting fluctuations.  Dark red and light blue lines indicate power-law models of 
	signal and noise distributions, respectively.  Dotted light blue lines indicate empirical estimates of
	the noise distribution from each of 3 disjoint analysis periods, showing that the background model 
	$p(\rho_c|\isbg) \propto \rho_c^{-54.8}$ is sufficiently accurate in the range of interest. 
	Dark green dashed lines show the total expected number of events (signal$+$noise) as a $90\%$ 
	credible band.   
	}
	\label{fig:er4_fgmc_snr}
\end{figure}

We first determine the rates of signal and noise events and the relative probabilities of signal vs.\ noise 
origin for each event~\cite{FGMC,O1BBH,GW150914:rates},
given only the $\rho_c$ value of each event, models of
the signal and noise event distributions over $\rho_c$, and an estimate of the total rate of noise events 
derived from time-shifted analyses~\cite{ihope,pycbc-software}.  
The result of this estimate is summarized in Fig.~\ref{fig:er4_fgmc_snr}. 
We find 53 events with a signal probability $p_{\mathrm{astro}}$ above 50\%, of which 47 have 
$p_{\mathrm{astro}}>90\%$.  This analysis is comparable to those used to estimate the rate and 
$p_{\mathrm{astro}}$ for binary black hole mergers in the first Advanced LIGO Observing period~\cite{O1BBH},
and does not use information about the mass distributions of signal or noise events, besides the 
assumption that the signal population is contained within the analysis mass limits. 

We now turn to our analysis, which estimates the rate and population model parameters simultaneously.  
The population model used here is a power law in each component mass, of the form
\begin{equation}
	p(m_1, m_2 | \popfg, \class = \mathrm{F}) \propto \begin{cases} m_1^{\alpha} m_2^{\beta} & \mathrm{if} \; 
 m_\mathrm{low} < m_2 \le m_1 < m_\mathrm{high} \\ 0 & \mathrm{else} \end{cases} \\
\end{equation}
where $\alpha$ and $\beta$ are the two power law slopes, both with true values equal to $0$. 
The mass cut-offs $m_\mathrm{low}$ and $m_\mathrm{high}$ are shared between both power laws, 
resulting in four free parameters in our population model. 
The selection effects were simulated numerically using the \texttt{LALsimulation}~\cite{LALsimulation} 
implementation of the IMRPhenomPv2 waveform~\cite{Hannam:2013oca,Khan:2015jqa} to implement the method 
described by Finn~\&~Chernoff~\cite{FinnChernoff93}.

As the background model, we used a power law fit to the distribution of $\rho_c$ values from the 
\texttt{pycbc} time-shifted analysis, giving a slope of $\approx-54.8$ for $\rho_c > 8$. 
In a separate step we constructed two dimensional fit to the distribution of component masses in the 
search results.  We find empirically that the background distribution can be approximated as the product of 
a power-law in chirp mass and a exponential distribution in mass ratio.
Using the masses as determined by the search is not strictly the correct approach, which would be to run the 
time-shifted data through the same LALInference analysis as used for the zero-lag events: this was, 
though, computationally infeasible.  Therefore the search results serve as a proxy for the optimal analysis. 
We have found empirically that small changes to the background mass distribution do not have a strong effect 
on the result, which is expected given the dissimilarity of foreground and background distributions. 
We do not include any additional uncertainty on the background model rate or mass distribution.
\begin{figure}
	\includegraphics[width=0.8\textwidth]{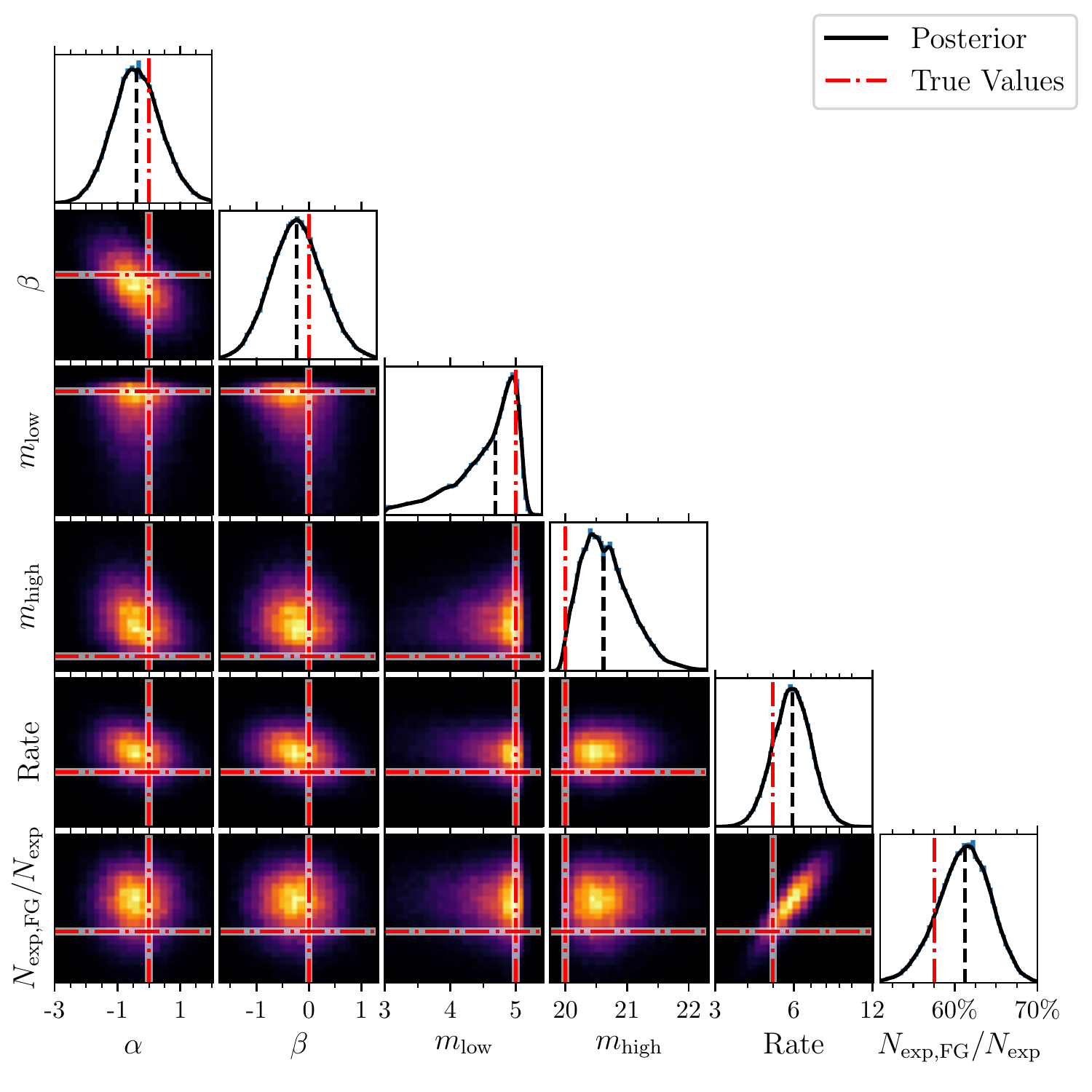}
	\centering
	\caption{Parameter estimates for the ER4 dataset.
			The foreground population model is power law in both component masses with separate 
			slopes $\alpha$ and $\beta$ and shared cut-offs $m_\mathrm{low}$ and $m_\mathrm{high}$.
			The cut-off masses and merger rate density are given in units of $\mathrm{M}_\odot$ and
			$\mathrm{Gpc}^{-3}\mathrm{yr}^{-1}$ respectively.
			The black lines show the kernel density estimate of the posterior, the thin blue lines the corresponding histogram.
			The red dash-dotted line indicates the true value.}
	\label{fig:er4_corner}
\end{figure}

The results of estimating the population parameters as displayed in figure~\ref{fig:er4_corner} 
show that we successfully recover the true population parameters. 
The slopes are underestimated slightly which causes the inferred merger rate density to be elevated, 
although it still encompasses the true value. 
The mass cut-offs are found well with some tails to lower or higher masses for $m_\mathrm{low}$ and 
$m_\mathrm{high}$ respectively, since the uniform distribution of injections covers all regions 
of the mass range without major gaps and the cut-offs are shared between both component masses.
\begin{figure}
	\includegraphics[width=0.7\textwidth]{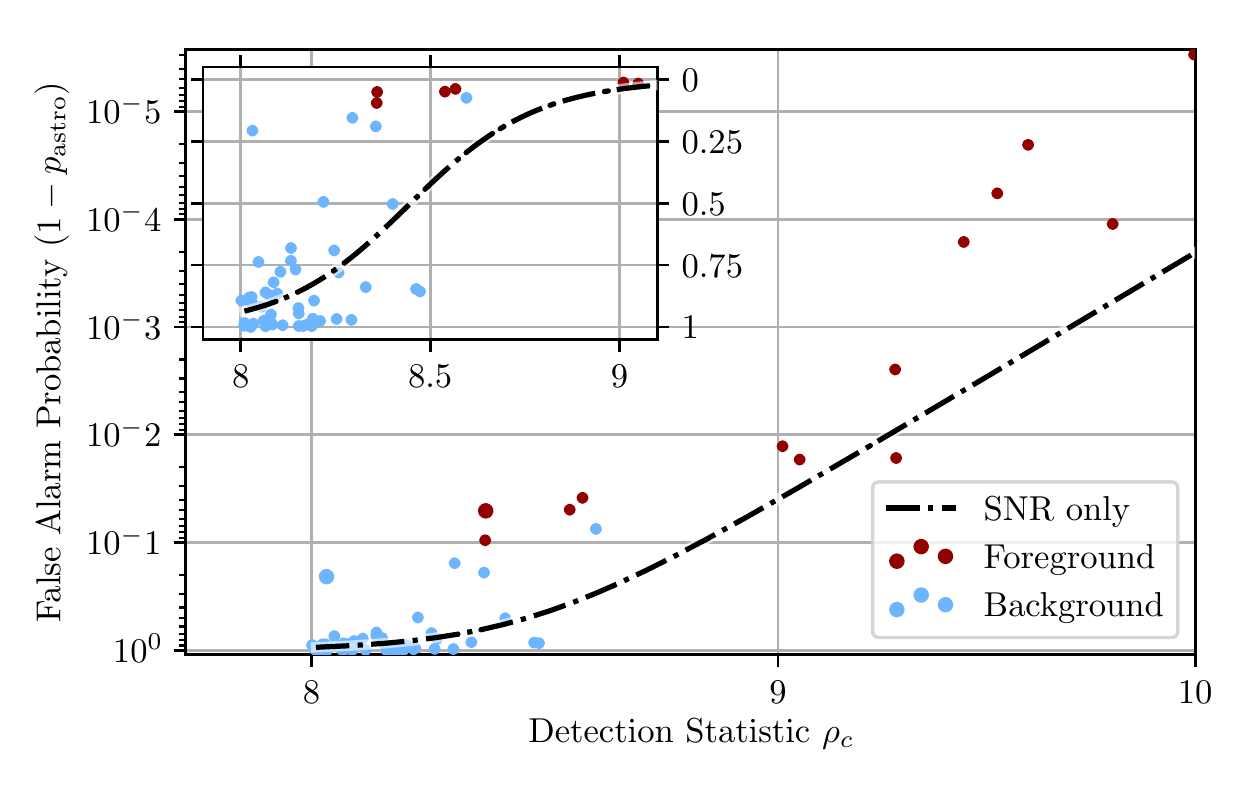}
	\centering
	\caption{The probability of an event recovered from the ER4 data being caused by noise 
			instead of corresponding to an astrophysical event, which is $1 - p_{\mathrm{astro}}$.
			Blue and red dots represent foreground and background events respectively.
			The dash-dotted line shows the probability that would be inferred by a SNR based estimate~\cite{FGMC}.
			The inset focuses on the region with background events and emphasises events 
			which are unlikely to be astrophysical by using a linear scale.}
	\label{fig:er4_fap}
\end{figure}

Since the slope of the background SNR distribution is much larger than the slope of the 
astrophysical foreground, the transition region where events may belong to either source category 
with comparable probabilities is quite small. This means there are few events which fall in between 
the region of certain background and certain foreground, limiting the gains that can be made by our 
method in this case. 
Nevertheless, the fact that the foreground mass distribution is quite distinct from the background 
causes a significant increase in $p_\mathrm{astro}$ relative to the $\rho_c$-based approach, as can be 
seen in figure~\ref{fig:er4_fap}. 

As with the previous ranking statistic based analysis we can count the events found with a 
$p_\mathrm{astro}$ value above some given threshold.
We find $56$ events above a threshold of $50\%$, which is 3 events more than before. 
The number of included events which we later identified as noise triggers rises from $4$ to $5$.
When the threshold is set to $p_\mathrm{astro} > 90\%$, the number of events increases by $5$ to $52$, 
though it now includes one noise trigger. 
Thus, we find that applying our new method to realistic data not only reproduces 
the results obtained using established methods, but identifies additional foreground events.
We also successfully recover the injected population parameters.

\section{Outlook}\label{sec:conclusion}
In this work we have derived a new technique for simultaneous estimation of parameters defining the shape of 
two or more sub-populations and their expected contribution to the overall number of events. 
This technique allows us to extract information from formerly sub-threshold events without biasing 
the result due to uncertainties in classifying their origin. The method is agnostic to the specific
choice of threshold, and lowering the threshold will allow more information to be included from events
of an uncertain nature, improving the result until the noise completely dominates the signals.

The greatest gains over existing methods are found when there is a large number of events for which the 
source classification gives comparable probabilities for at least two categories, 
while the distributions in secondary parameters are very distinct.
This behaviour near the transition between populations is likely to be especially useful in the 
characterisation of weak event populations, such as unresolvable binary mergers at cosmological distances. 
This is of particular interest when determining whether the source population evolves with redshift.
However, this also implies that gains are expected to be small when the primary source classification is very 
potent and population models are uncertain; in this case our method converges to that with a single population. 

While such thresholded analyses that ignore the possible presence of background cannot be guaranteed free
of systematic bias, the expected size of bias can be bounded by considering the rate of background events
above threshold, as well as the degree of divergence between foreground and background
distributions over the parameters of interest.  Controlling the bias of a thresholded analysis thus still 
requires accurate background estimation.  In particular, for the small number of events thus far 
detected by LIGO-Virgo, possible biases on population inference due to neglecting background contamination 
are expected to be well below statistical errors.  

We have successfully tested this new model on different binary merger mass distributions in an artificial 
universe, as well as to synthetic LIGO data from an engineering run.
This demonstrates the feasibility of applying this method to existing and future LIGO-Virgo observing runs, which
should allow a better joint determination of source event rates and distributions.
The method itself is, however, not limited to the realm of gravitational wave astronomy, and can be useful 
whenever a set of data points contains multiple populations.

\section*{Acknowledgements}
We are grateful to the LSC for allowing the use of LIGO software engineering run data hosted on the 
CIT cluster, and access to LIGO Data Grid computing resources. 
We are grateful to Chris Pankow for useful discussions and assistance in the ER4 analysis. 
We would also like to thank Jonathan Gair and Richard O'Shaughnessy for their helpful comments and feedback.
This work was supported by STFC grants ST/K005014/2 and ST/M004090/2.

\bibliography{references}

\end{document}